\documentclass[aps,prl,amsmath,amssymb,twocolumn,preprintnumbers,nofootinbib]{revtex4}

\usepackage[english]{babel}
\usepackage[hidelinks]{hyperref}
\usepackage{graphicx}
\usepackage[utf8]{inputenc}

\renewcommand{\d}{\textrm{d}}

\newcommand{\vac}{\textrm{vac}}
\newcommand{\W}{\mathcal{W}}
\newcommand{\N}{\mathcal{N}}
\newcommand{\I}{\mathcal{I}}
\newcommand{\A}{\mathcal{A}}
\renewcommand{\H}{\mathcal{H}}
\renewcommand{\L}{\mathcal{L}}
\renewcommand{\O}{\mathcal{O}}
\newcommand{\hc}{\textrm{h.c.}}
\newcommand{\alphadot}{{\dot{\alpha}}}
\renewcommand{\bar}[1]{\overline{#1}}
\DeclareMathOperator{\tr}{tr}

\begin{document}

\title{The local renormalization of super-Yang-Mills theories}
\author{Marc Gillioz}
\affiliation{UC Davis Department of Physics, \\
One Shields Avenue, Davis, CA 95616}
\date{\today}

\begin{abstract}

We show how to consistently renormalize $\N = 1$ and $\N = 2$ super-Yang-Mills theories in flat space with a local (i.e.~space-time-dependent) renormalization scale in a holomorphic scheme. The action gets enhanced by a term proportional to derivatives of the holomorphic coupling. In the $\N = 2$ case, this new action is exact at all orders in perturbation theory.

\end{abstract}

\maketitle

\section{Introduction}

It is customary when performing the renormalization of quantum field theories to introduce a constant renormalization scale, required by dimensional analysis. In the Wilsonian approach, renormalized quantum field theories can be viewed as effective field theories valid below a specific cutoff, and the bare couplings are adjusted in a cutoff-dependent way so that the low-energy theory is actually invariant under an infinitesimal global deformation of the cutoff. This approach, however, 
is not the most general: in principle, nothing prohibits us from deforming the cutoff in a space-time-dependent manner, meaning that we can choose a different renormalization scale at different points in time and space.
As long as the deformations of the cutoff are smooth, the renormalized theory should still make perfect sense.
In this way, one introduces a space-time dependence in the couplings of the theory, meaning that the coupling constants are promoted to (background) fields. Besides modifying the equations of motion for the quantum fields of the theory, the space-time-dependent couplings also act as external sources for composite operators. 
For these composite operators to be renormalized too, new counterterms have to be introduced in the theory. These counterterms have physical significance, as we shall see below.
The ideas summarized so far are found in the literature under the name of \emph{local renormalization group}~\cite{Drummond:1977dg, Shore:1986hk, Osborn:1987au, Osborn:1989td, Jack:1990eb, Osborn:1991gm}%
\footnote{Recent reviews of these old ideas by the original authors can be found in Refs.~\cite{Jack:2013sha, Shore:2016xor}.}.

One aspect of the local renormalization group (RG) has been studied in particular so far: when coupled to a classical background metric, the local RG allows to write a generalization of the Callan-Symanzik equation that describes the response of the theory under a local Weyl rescaling of the metric, $g^{\mu\nu} \to \Omega(x)^2 g^{\mu\nu}$. The local RG equation describes the properties of correlators involving the energy-momentum tensor. It was used to provide an independent derivation of Zamolodchikov $c$-theorem in two dimensions~\cite{Zamolodchikov:1986gt}, and to derive an analogous $a$-theorem in four dimensions, proving the irreversibility of the RG flow at leading order in perturbation theory~\cite{Osborn:1989td, Jack:1990eb, Osborn:1991gm, Baume:2014rla,Keren-Zur:2014sva}.
The local RG equation also played a crucial role in elucidating the relation between scale invariance and conformal invariance in four dimensions~\cite{Luty:2012ww, Fortin:2012hn, Nakayama:2013is}. It was used as well to try and generalize the $a$-theorem in dimensions other than two and four~\cite{Baume:2013ika, Grinstein:2014xba, Osborn:2015rna, Gracey:2015fia}.
A superspace version of the local RG equation for supersymmetric theories has been written down~\cite{Auzzi:2015yia}, and the $a$-function obtained in this way has been shown to agree with non-perturbative derivations using $R$-symmetry anomalies~\cite{Freedman:1998rd} or the dilaton effective action in a curved background~\cite{Prochazka:2015edz}. The same method allowed for new constraints in the study of supersymmetric conformal manifolds~\cite{Gomis:2015yaa}.

Most of the achievements of the local RG so far involve renormalization in curved space-time. It is convenient to couple a locally renormalized theory to a background metric, as it makes the local rescalings explicitly realized as Weyl transformations. However, it is by no means necessary. This work explores the consequences of the local renormalization group in a pure flat space approach. As mentioned above, the renormalizability of composite operators require to augment the action with additional terms. These are terms that do not involve the quantum fields but only the couplings, or more precisely derivatives thereof. They can be computed order by order in perturbation theory, and are completely determined by the field content of the theory and the choice of renormalization scheme. In this work, we focus on supersymmetric gauge theories, for which there is a preferred scheme, the holomorphic scheme, and powerful non-renormalization theorems.

Our results for both $\N=1$ and $\N=2$ theories can be summarized in one equation: the action of a renormalized supersymmetric gauge theory must be enhanced by a new ``vacuum'' contribution
\begin{equation}
	S_\vac = 
	\frac{d_G}{T_G}
	\int \d^4 x\d^4\theta \, \left[
	\frac{(\partial_\mu \tau) (\partial^\mu \bar{\tau})}
	{8 \, b}
	+ \frac{(D\tau)^2 (\bar{D} \bar{\tau})^2}
	{192 \, b^3} \right]
	\label{eq:Svac}
\end{equation}
where $\tau$ is the holomorphic coupling promoted here to a chiral superfield, $d_G$ is the dimension of the gauge group, $T_G$ the Dynkin index of its adjoint representation (e.g.~$d_G = N_c^2 - 1$ and $T_G = N_c$ for a $SU(N_c)$ gauge theory), and $b$ is the one-loop coefficient of the $\beta$-function, i.e.~$b = 3 T_G/(8\pi^2)$ for $\N = 1$ and $b = 2 T_G/(8\pi^2)$ for $\N = 2$. We describe in the next sections how this result is obtained from a one-loop computation, and how it remains valid at all orders for $\N = 2$.

\section[N=2 with a space-time-dependent cutoff]{$\N = 2$ with a space-time-dependent cutoff}

The idea of this work is to use the Wilsonian approach to define the $\N=2$ (and later $\N=1$) super-Yang-Mills theory in terms of a finite, $\N = 4$ theory softly broken by a space-time-dependent mass term for the matter hypermultiplet. We follow in particular the derivation of Ref.~\cite{ArkaniHamed:1997mj}. The $\N = 4$ theory in isolation  is finite, and therefore its gauge coupling $g_0$ and vacuum angle $\Theta$ are not renormalized. We make use of $\N=1$ superspace formulation throughout the derivation, in which the $\N = 4$ Lagrangian can be written
\begin{eqnarray}
	\L_{\N=4} & = & \frac{1}{8} \int \d^2\theta
	\left( \frac{1}{g_0^2} + \frac{i \Theta}{8 \pi^2} \right)
	\tr(\W^\alpha \W_\alpha) + \hc
	\nonumber \\
	&& + \int \d^4\theta \frac{2}{g_0^2} \tr\left(
	\bar{\chi} e^{-2V} \chi e^{2V} \right)
	\nonumber \\
	&& + \sum_{i=1}^2 \int \d^4\theta \frac{2}{g_0^2} \tr\left(
	\bar{\Phi}_i e^{-2V} \Phi_i e^{2V} \right)
	\nonumber \\
	&& + \int \d^2\theta \frac{\sqrt{2}}{g_0^2}
	\tr\left( \chi [ \Phi_1, \Phi_2 ]\right) + \hc
	\label{eq:sYM:4}
\end{eqnarray}
There are three chiral superfields, $\chi$, $\Phi_1$ and $\Phi_2$, a vector field $V$, and $\W^\alpha$ is the field strength tensor associated to $V$. In $\N = 2$ language, $(\chi, V)$ forms a gauge supermultiplet and $(\Phi_1, \Phi_2)$ a matter hypermultiplet. We want to softly deform the theory by adding a mass term for the $\Phi_i$, so that a pure $\N=2$ gauge theory is recovered below that mass scale. Before proceeding, however, we perform a rescaling of the hypermultiplet as $\Phi_i \to g_0 \Phi_i$. The Jacobian of this transformation is non-trivial, and we get therefore
\begin{eqnarray}
	\L_{\N=4} & = & \frac{1}{8} \int \d^2\theta
	\left( \frac{1}{g_c^2} + \frac{i \Theta}{8 \pi^2} \right)
	\tr(\W^\alpha \W_\alpha) + \hc
	\nonumber \\
	&& + \int \d^4\theta \frac{2}{g_c^2} \tr\left(
	\bar{\chi} e^{-2V} \chi e^{2V} \right)
	\nonumber \\
	&& + \sum_{i=1}^2 2 \int \d^4\theta \tr\left(
	\bar{\Phi}_i e^{-2V} \Phi_i e^{2V} \right)
	\nonumber \\
	&& + \sqrt{2} \int \d^2\theta\,
	\tr\left( \chi [ \Phi_1, \Phi_2 ]\right) + \hc
	\label{eq:sYM:4:canonical}
\end{eqnarray}
where we have now
\begin{equation}
	\frac{1}{g_c^2} = \frac{1}{g_0^2}
	+ \frac{2 T_G}{8\pi^2} \log(g_0).
	\label{eq:canonicalcoupling:2}
\end{equation}
Eq.~\eqref{eq:sYM:4:canonical} is a fancy way of describing $\N=4$ super-Yang-Mills, where the supersymmetries are not obvious. The finiteness of the theory is nevertheless preserved, and the hypermultiplet fields $\Phi_i$ are now canonically normalized and ready to be integrated out.

We deform this theory by adding a soft mass term for the hypermultiplet $(\Phi_1, \Phi_2)$,
\begin{equation}
	\delta \L = \sum_{i=1}^2 \int \d^2\theta \,
	\Lambda \tr( \Phi_i \Phi_i ) + \hc
\end{equation}
where $\Lambda$ is taken to be an external chiral multiplet, whose vacuum expectation value defines a mass scale $M$:
\begin{equation}
	\langle \Lambda \rangle
	= \langle \bar{\Lambda} \rangle = M.
\end{equation}
In the Wilsonian spirit, we have to replace the bare coupling $g_0$ by a functional of $\Lambda$ in such a way that the low-energy theory at energies $p^2 \ll M^2$ remains invariant under an infinitesimal variation of $ \Lambda$. This can be done by considering the Lagrangian
\begin{eqnarray}
	\L_{\N=2} & = & \frac{1}{8} \int \d^2\theta \,
	\tau[\Lambda] \tr (\W^\alpha \W_\alpha) + \hc
	\nonumber \\
	&& + \int \d^4\theta \left( \tau[\Lambda] +
	\bar{\tau} [\bar{\Lambda}] \right) \tr\left(
	\bar{\chi} e^{-2V} \chi e^{2V} \right)
	\nonumber \\
	&& + \sum_{i=1}^2 2 \int \d^4\theta \tr\left(
	\bar{\Phi}_i e^{-2V} \Phi_i e^{2V} \right)
	\nonumber \\
	&& + \sum_{i=1}^2 \int \d^2\theta \,
	\Lambda \tr( \Phi_i \Phi_i ) + \hc
	\nonumber \\
	&& + \sqrt{2} \int \d^2\theta\,
	\tr\left( \chi [ \Phi_1, \Phi_2 ]\right) + \hc
	\nonumber \\
	&& + \L_\vac[\Lambda, \bar{\Lambda}].
	\label{eq:sYM:2}
\end{eqnarray}%
\begin{figure}
	\centering
	\parbox{2.5cm}{\includegraphics{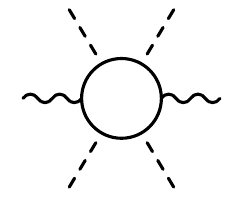} \\ (a)}
	\parbox{2.5cm}{\includegraphics{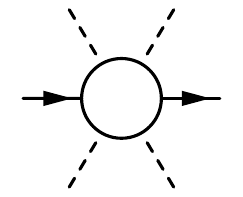} \\ (b)}
	\parbox{2.5cm}{\includegraphics{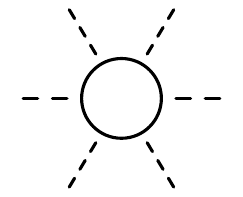} \\ (c)}
	\caption{A sample of the superspace  diagrams contributing to the effective action of $\N=2$ super-Yang-Mills. The fields $\Phi_i$ are here running in the loop. The external legs consist in gauge superfields (a), chiral matter fields (b) or the chiral field $\Lambda$ (a,b and c).}
	\label{fig:diagrams:2}
\end{figure}%
$\tau[\Lambda]$ is fixed by requiring that the theory does not depend on $\Lambda$ after integrating out the $\Phi_i$ superfields. It can be determined in practice by writing $\Lambda = M + \delta\Lambda$ and performing a perturbative expansion, considering $\delta\Lambda \tr(\Phi_i \Phi_i)$ as an interaction term in the theory of two massive chiral superfields. The only relevant Feynman diagrams are those with a loop of $\Phi_1$ or $\Phi_2$ and external legs consisting in $V$, $\chi$ and $\delta\Lambda$, as illustrated in Fig.~\ref{fig:diagrams:2}. The low-energy effective theory is then obtained by only considering terms of order $M^n$ with $n \geq 0$. By simple power counting and taking into account gauge invariance, diagrams with $V$ or $\chi$ external legs cannot have $\delta\Lambda$-legs that carry momentum. The low-energy action in the gauge and matter sectors is that of a theory with constant $\Lambda$, and we find therefore
\begin{equation}
	\tau[\Lambda] = \frac{1}{g_0^2}
	+ \frac{i \Theta}{8 \pi^2}
	+ \frac{2 T_G}{8\pi^2}
	\log\left( \frac{\Lambda}{\mu} \right).
	\label{eq:running:2}
\end{equation}
This is the ordinary global renormalization group result, obtained in the local renormalization framework. Notice that the $\log(g_0)$ term in eq.~\eqref{eq:canonicalcoupling:2} could be canceled by a redefinition $\Lambda \to g_0 \Lambda$.

There is however a crucial difference between local and global renormalization: in our method, there are infinitely many more diagrams of order $M^0$ which do not decouple in the infrared: these are all diagrams with $\delta\Lambda$ external legs only.
Their contribution must be canceled by the term denoted $\L_\vac$ in the action~\eqref{eq:sYM:2}. Only diagrams with at most two powers of momentum on the external legs can contribute in the regime $p^2 \ll M^2$, and diagrams with less than two powers of momentum actually vanish identically. Each of the relevant diagrams is separately free of ultraviolet divergences from $\N = 4$ supersymmetry, but summing all of them is a rather advanced combinatorics problem. Instead, we construct the propagator for $\Phi$ with a space-time-dependent mass in a derivative expansion, and use the relation between the propagator and the integrand to evaluate the sum of diagrams, as explained in the appendix. The result is finite and independent of $M$:
\begin{equation}
	\L_\vac = \frac{d_G}{(4 \pi)^2}
	\int \d^4\theta \left[
	\frac{\partial_\mu \Lambda \partial^\mu \bar{\Lambda}}
	{2\, \Lambda \bar{\Lambda}}
	+ \frac{D^\alpha \Lambda D_\alpha \Lambda
	\bar{D}_\alphadot \bar{\Lambda}
	\bar{D}^\alphadot \bar{\Lambda}}
	{48 \, \Lambda^2 \bar{\Lambda}^2} \right],
	\label{eq:Lvac:Lambda}
\end{equation}
where $D_\alpha$ and $\bar{D}_\alphadot$ are the SUSY-covariant derivatives. Notice that this Lagrangian only depend on the logarithm of $\Lambda$ and $\bar{\Lambda}$, and as such it can be rewritten in terms of the holomorphic coupling $\tau$ using the relation~\eqref{eq:running:2},
\begin{equation}
	\L_\vac = \frac{d_G}{(4\pi)^2}
	\int \d^4\theta \left[ \frac{1}{2 \, b^2}
	(\partial \tau)^2
	+ \frac{1}{48 \, b^4} (D \tau)^2
	(\bar{D} \bar{\tau})^2 \right],
	\label{eq:Lvac:2}
\end{equation}
where $b$ is the $\beta$-function coefficient defined in the introduction.
This form makes it explicit that $\L_\vac$ remains part of the $\N = 2$ super-Yang-Mills theory when we take the limit $\Lambda \to \infty$. It should also be emphasized that $\L_\vac$ is exact at all loop orders in perturbation theory: since $\Phi$ in the theory~\eqref{eq:sYM:2} does not have self-interactions, there are literally no diagrams contributing to $\L_\vac$ beyond the one loop order.

\section[Extension to N = 1]{Extension to $\N = 1$}

The $\N=1$ theory can be derived along the same lines, with substantially similar results. The important difference is that our result is valid at one-loop only, higher-order corrections being important in principle.

The derivation of the $\N = 1$ action from $\N = 4$ proceeds as in the previous section, except that we now introduce a soft mass term for all three chiral multiplets, $\Phi_1$, $\Phi_2$ and $\chi$ (which we rename $\Phi_3$ for simplicity). The Lagrangian is therefore
\begin{eqnarray}
	\L_{\N=1} & = & \frac{1}{8} \int \d^2\theta \,
	\tau[\Lambda] \tr (\W^\alpha \W_\alpha) + \hc
	\nonumber \\
	&& + \sum_{i=1}^3 2 \int \d^4\theta \tr\left(
	\bar{\Phi}_i e^{-2V} \Phi_i e^{2V} \right)
	\nonumber \\
	&& + \sum_{i=1}^3 \int \d^2\theta \,
	\Lambda \tr( \Phi_i \Phi_i ) + \hc
	\nonumber \\
	&& + \sqrt{2} g_0 \sum_{i,j,k=1}^3
	\frac{\epsilon_{ijk}}{3}\int \d^2\theta\,
	\tr\left( \Phi_i \Phi_j \Phi_k \right) + \hc
	\nonumber \\
	&& + \L_\vac[\Lambda, \bar{\Lambda}],
	\label{eq:sYM:1}
\end{eqnarray}%
\begin{figure}
	\centering
	\parbox{2.5cm}{\includegraphics{diag-1} \\ (a)}
	\parbox{2.5cm}{\includegraphics{diag-3} \\ (b)}
	\parbox{2.5cm}{\includegraphics{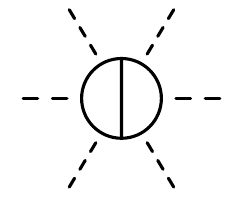} \\ (c)}
	\caption{Same as Fig.~\ref{fig:diagrams:2} for $\N = 1$ super-Yang-Mills. This time there are no external chiral matter superfield, but multi-loop diagrams (c) contribute to $\L_\vac$.}
	\label{fig:diagrams:1}
\end{figure}
where we must now require
\begin{equation}
	\tau[\Lambda] =
	\frac{1}{g_0^2}
	+ \frac{i \Theta}{8 \pi^2}
	+ \frac{3 T_G}{8 \pi^2}
	\log\left( \frac{\Lambda}{\mu} \right)
	+ \O(g_0^2).
	\label{eq:running:1}
\end{equation}
Not surprisingly, the cutoff dependence of $\tau$ agrees again with the global renormalization group approach.
As before, we have rescaled $\Lambda \to g_0 \Lambda$ to get rid of the $\log(g_0)$ appearing in the definition of the gauge coupling for canonically normalized matter fields. While this rescaling was absolutely insignificant in the $\N = 2$ case, for $\N = 1$ it is absolutely necessary for $\tau$ to make sense as a perturbative expansion in powers of the bare coupling $g_0$. In other words, for $\N = 1$ theories the vacuum contribution $\L_\vac$ depends on the choice of renormalization scheme, and our results apply to the holomorphic scheme only. The computation of $\L_\vac$ in this case goes exactly as before, the only difference being that there are now three different chiral fields running in the loop, so that we recover eq.~\eqref{eq:Lvac:Lambda} multiplied by a factor of $\frac{3}{2}$, and thus
\begin{eqnarray}
	\L_\vac & = & \frac{d_G}{(4\pi)^2}
	\int \d^4\theta \left[ \frac{3}{4 \, b^2}
	(\partial \tau)^2
	+ \frac{1}{32 \, b^4} (D \tau)^2
	(\bar{D} \bar{\tau})^2 \right],
	\nonumber \\
	&& + \textrm{higher-loop corrections.}
	\label{eq:Lvac}
\end{eqnarray}
The investigation of higher-order corrections are beyond the scope of this work; notice that a supersymmetry-breaking anomaly has been reported to appear at the two-loop order~\cite{Kraus:2001tg, Kraus:2001id, Kraus:2002nu, Babington:2005vu}.%

\section{Conclusions}

We have shown in this work how the local renormalization group can be derived in a Wilsonian approach to $\N = 2$ and $\N = 1$ super-Yang-Mills theories. The trick of promoting the couplings of supersymmetric theories to background sources has been used before, time-honored examples being the Shifman-Vainshtein derivation of the all-order $\beta$-function for supersymmetric theories with matter~\cite{Shifman:1986zi, Shifman:1991dz} or the study of soft supersymmetry breaking effects relevant for phenomenology~\cite{Giudice:1997ni, ArkaniHamed:1998kj}.
But to the best of our knowledge, it is the first time that the presence of the ``vacuum'' term~\eqref{eq:Svac} in the effective action is emphasized. Such a term is reminiscent of the dilaton effective action used in the non-perturbative proof of the \emph{a} theorem~\cite{Komargodski:2011vj, Komargodski:2011xv}, and its variation with respect to the holomorphic coupling is indeed related to the conformal anomalies through the work of Jack and Osborn~\cite{Jack:1990eb, Osborn:1991gm, Jack:2013sha, Auzzi:2015yia}. The $a$ function must in particular obey
\begin{equation}
	\frac{\d a}{\d \tau}
	\propto \beta(\tau)
	\left. \frac{\delta^2 S}{\delta \tau \delta \bar{\tau}}
	\right|_{\tau = \textrm{const}}
	\propto \frac{d_G}{T_G},
	\label{eq:atheorem}
\end{equation}
The exact form of this relation and its consequences will be determined elsewhere~\cite{Gillioz:xxx}, as the details of this equality have only been worked out for non-supersymmetric field theories. For the same reason we cannot directly compare our result to existing computations~\cite{Osborn:2003vk, Nakayama:2015ita}. Nevertheless, it can be noted already that the monotonicity of $a$ at all orders in perturbation theory can be probably be made obvious for $\N = 2$. Eq.~\eqref{eq:atheorem} also indicates that non-perturbative effects must become relevant along the renormalization flow to prevent $a$ from running all the way to negative values~\cite{Antipin:2013qya, Antipin:2013pya}.

The study of non-perturbative phenomena like instantons and monopoles in the local RG framework is also of foremost interest. The novelty there is that $\L_\vac$  arises as a Lagrangian term and is well-defined all along the renormalization group flow, and can therefore be incorporated in semi-classical computations, in which one can imagine allowing the renormalization scale to vary over space and/or time~\cite{Gillioz:xxx}. A true semi-classical result should be independent of the choice of renormalization scale, even locally. 
Other directions deserve further studies as well, as for instance the extension of our results to models with matter fields, or the investigation of the higher-loop structure of $\L_\vac$.

\subsection*{Acknowledgments}

The author would like to thank Markus Luty for useful discussions, as well as the people at CP$^3$--Origins where this project was initiated a long time ago.
This work is supported by the Swiss National Science Foundation (SNSF) grant number P300P2154559.

\appendix
\section{Appendix: Computation of the effective action}

In this appendix, we provide more details on the derivation of eq.~\eqref{eq:Lvac:Lambda}. The first step consists in evaluating the propagator for the chiral superfield $\Phi$ in the presence of a space-time dependent mass term $\Lambda$ in a derivative expansion. We can write the Lagrangian for $\Phi$ as
\begin{equation}
	\mathcal{L}_\Phi = \int \d^4\theta \, \frac{1}{2}
	\left( \Phi ~~ \bar{\Phi} \right) \Omega
	\left( \begin{array}{c}
		\Phi \\ \bar{\Phi}
	\end{array}\right),
\end{equation}
where we defined,
\begin{equation}
	\Omega = \left(\begin{array}{cc}
		-\Lambda \frac{D^2}{4 \square} & 1 \\
		1 & -\bar{\Lambda} \frac{\bar{D}^2}{4 \square}
	\end{array}\right),
\end{equation}
making use of the (non-local) chiral projectors $\frac{D^2}{4 \square}$, $\frac{\bar{D}^2}{4 \square}$.
The propagator for the doublet $( \Phi, \bar{\Phi} )$ is then proportional to the inverse of the matrix $\Omega$,
\begin{equation}
	\Delta = \H 
	\left(\begin{array}{cc}
		\bar{\Lambda} \frac{D^2}{4\square}
		& -\frac{D^2 \bar{D}^2}{16 \square} \\
		-\frac{\bar{D}^2 D^2}{16 \square}
		& \Lambda \frac{\bar{D}^2}{4\square}
	\end{array}\right),
\end{equation}
where $\H$ is a derivative operator satisfying
\begin{equation}
	\H \cdot \left( \square + \Lambda \bar{\Lambda}  \right)
	= 1.
\end{equation}
$\H$ can be computed term-by-term in a derivative expansion:
\begin{equation}
	\H = \frac{1}{\square + \Lambda \bar{\Lambda}}
	+ 2 [\partial_\mu (\Lambda \bar{\Lambda})]
	\frac{1}{[\square + \Lambda \bar{\Lambda}]^3}
	\partial^\mu + \O(\partial^2 \Lambda \bar{\Lambda}),
\end{equation}
where we used the notation
\begin{equation}
	\frac{1}{[\square + \Lambda \bar{\Lambda} ]^n}
	\equiv \sum_{k=0}^\infty (-1)^k
	\frac{(n+k-1)!}{(n-1)! k!}
	\frac{1}{(\Lambda \bar{\Lambda})^{n+k}} \square^k.
\end{equation}
In the limit $\Lambda, \bar{\Lambda} \to M$, we recover the usual propagator for a chiral superfield of mass $M$,
\begin{equation}
	\Delta_0 = \frac{1}{\square + M^2}
	\left(\begin{array}{cc}
		M \frac{D^2}{4\square}
		& -\frac{D^2 \bar{D}^2}{16 \square} \\
		-\frac{\bar{D}^2 D^2}{16 \square}
		& M \frac{\bar{D}^2}{4\square}
	\end{array}\right).
\end{equation}
By writing down the full propagator $\Delta$ in the presence of background sources, we have actually summed over arbitrarily many insertions of the chiral superfields $\Lambda$ and $\bar{\Lambda}$. The result can be expressed as a single interaction term $\I$, obeying
\begin{equation}
	\Delta = \Delta_0 + \Delta_0 \cdot \I \cdot \Delta_0.
\end{equation}
The exact form of the matrix $\I$ as a derivative expansion can be resolved with the help of computer algebra. Its leading term turns out to be of order $\O(D^2\Lambda)$, where $D$ indicates the SUSY-covariant derivative acting on an external source $\Lambda$ (or $\bar{D}$ acting on $\bar{\Lambda}$). The sum of Feynman diagrams
\begin{equation}
	\A \equiv \raisebox{-0.625cm}{\includegraphics{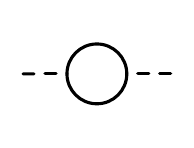}}
	+ \raisebox{-0.625cm}{\includegraphics{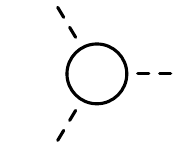}}
	+ \raisebox{-0.625cm}{\includegraphics{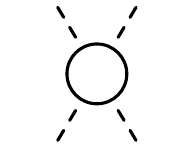}}
	+ \ldots
\end{equation}
is then readily given in terms of $\I$ by
\begin{equation}
	\A = \tr(\Delta_0 \cdot \I) + \frac{1}{2}
	\tr(\Delta_0 \cdot \I \cdot \Delta_0 \cdot \I) + \ldots,
\end{equation}
where higher-order terms can be neglected as they are at least of order $\O(D^6\Lambda)$ and their contribution to the low-energy effective action is therefore suppressed by powers of $M$. Upon loop integration, terms with less than four SUSY-covariant derivatives vanish and we are left with two terms of order $\O(D^4 \Lambda)$ that have been reported in eq.~\eqref{eq:Lvac:Lambda}.

\bibliography{Bibliography}

\end{document}